\documentstyle[12pt]{article}

\expandafter\ifx\csname amssym.def\endcsname\relax \else\endinput\fi
\expandafter\edef\csname amssym.def\endcsname{%
       \catcode`\noexpand\@=\the\catcode`\@\space}
\catcode`\@=11
\def\undefine#1{\let#1\undefined}
\def\newsymbol#1#2#3#4#5{\let\next@\relax
 \ifnum#2=\@ne\let\next@\msafam@\else
 \ifnum#2=\tw@\let\next@\msbfam@\fi\fi
 \mathchardef#1="#3\next@#4#5}
\def\mathhexbox@#1#2#3{\relax
 \ifmmode\mathpalette{}{\m@th\mathchar"#1#2#3}%
 \else\leavevmode\hbox{$\m@th\mathchar"#1#2#3$}\fi}
\def\hexnumber@#1{\ifcase#1 0\or 1\or 2\or 3\or 4\or 5\or 6\or 7\or 8\or
 9\or A\or B\or C\or D\or E\or F\fi}

\ifcase\@ptsize
  \font\tenmsa=msam10
  \font\sevenmsa=msam7
  \font\fivemsa=msam5
\or
  \font\tenmsa=msam10  scaled \magstephalf
  \font\sevenmsa=msam7 scaled \magstephalf
  \font\fivemsa=msam5  scaled \magstephalf
\or
  \font\tenmsa=msam10  scaled \magstep1
  \font\sevenmsa=msam7 scaled \magstep1
  \font\fivemsa=msam5  scaled \magstep1
\fi

\newfam\msafam
\textfont\msafam=\tenmsa
\scriptfont\msafam=\sevenmsa
\scriptscriptfont\msafam=\fivemsa
\edef\msafam@{\hexnumber@\msafam}
\mathchardef\dabar@"0\msafam@39
\def\dashrightarrow{\mathrel{\dabar@\dabar@\mathchar"0\msafam@4B}}
\def\dashleftarrow{\mathrel{\mathchar"0\msafam@4C\dabar@\dabar@}}

\def\ulcorner{\delimiter"4\msafam@70\msafam@70 }
\def\urcorner{\delimiter"5\msafam@71\msafam@71 }
\def\llcorner{\delimiter"4\msafam@78\msafam@78 }
\def\lrcorner{\delimiter"5\msafam@79\msafam@79 }
\def\yen{{\mathhexbox@\msafam@55 }}
\def\checkmark{{\mathhexbox@\msafam@58 }}
\def\circledR{{\mathhexbox@\msafam@72 }}
\def\maltese{{\mathhexbox@\msafam@7A }}

\ifcase\@ptsize
  \font\tenmsb=msbm10
  \font\sevenmsb=msbm7
  \font\fivemsb=msbm5
\or
  \font\tenmsb=msbm10  scaled \magstephalf
  \font\sevenmsb=msbm7 scaled \magstephalf
  \font\fivemsb=msbm5  scaled \magstephalf
\or
  \font\tenmsb=msbm10  scaled \magstep1
  \font\sevenmsb=msbm7 scaled \magstep1
  \font\fivemsb=msbm5  scaled \magstep1
\fi

\newfam\msbfam
\textfont\msbfam=\tenmsb
\scriptfont\msbfam=\sevenmsb
\scriptscriptfont\msbfam=\fivemsb
\edef\msbfam@{\hexnumber@\msbfam}
\def\Bbb#1{{\fam\msbfam\relax#1}}
\def\widehat#1{\setbox\z@\hbox{$\m@th#1$}%
 \ifdim\wd\z@>\tw@ em\mathaccent"0\msbfam@5B{#1}%
 \else\mathaccent"0362{#1}\fi}
\def\widetilde#1{\setbox\z@\hbox{$\m@th#1$}%
 \ifdim\wd\z@>\tw@ em\mathaccent"0\msbfam@5D{#1}%
 \else\mathaccent"0365{#1}\fi}

\ifcase\@ptsize
  \font\teneufm=eufm10
  \font\seveneufm=eufm7
  \font\fiveeufm=eufm5
\or
  \font\teneufm=eufm10   scaled \magstephalf
  \font\seveneufm=eufm7  scaled \magstephalf
  \font\fiveeufm=eufm5   scaled \magstephalf
\or
  \font\teneufm=eufm10   scaled \magstep1
  \font\seveneufm=eufm7  scaled \magstep1
  \font\fiveeufm=eufm5   scaled \magstep1
\fi

\newfam\eufmfam
\textfont\eufmfam=\teneufm
\scriptfont\eufmfam=\seveneufm
\scriptscriptfont\eufmfam=\fiveeufm
\def\frak#1{{\fam\eufmfam\relax#1}}

\csname amssym.def\endcsname

\expandafter\ifx\csname pre amssym.tex at\endcsname\relax \else \endinput\fi
\expandafter\chardef\csname pre amssym.tex at\endcsname=\the\catcode`\@
\catcode`\@=11
\newsymbol\boxdot 1200
\newsymbol\boxplus 1201
\newsymbol\boxtimes 1202
\newsymbol\square 1003
\newsymbol\blacksquare 1004
\newsymbol\centerdot 1205
\newsymbol\lozenge 1006
\newsymbol\blacklozenge 1007
\newsymbol\circlearrowright 1308
\newsymbol\circlearrowleft 1309
\undefine\rightleftharpoons
\newsymbol\rightleftharpoons 130A
\newsymbol\leftrightharpoons 130B
\newsymbol\boxminus 120C
\newsymbol\Vdash 130D
\newsymbol\Vvdash 130E
\newsymbol\vDash 130F
\newsymbol\twoheadrightarrow 1310
\newsymbol\twoheadleftarrow 1311
\newsymbol\leftleftarrows 1312
\newsymbol\rightrightarrows 1313
\newsymbol\upuparrows 1314
\newsymbol\downdownarrows 1315
\newsymbol\upharpoonright 1316
 
\newsymbol\downharpoonright 1317
\newsymbol\upharpoonleft 1318
\newsymbol\downharpoonleft 1319
\newsymbol\rightarrowtail 131A
\newsymbol\leftarrowtail 131B
\newsymbol\leftrightarrows 131C
\newsymbol\rightleftarrows 131D
\newsymbol\Lsh 131E
\newsymbol\Rsh 131F
\newsymbol\rightsquigarrow 1320
\newsymbol\leftrightsquigarrow 1321
\newsymbol\looparrowleft 1322
\newsymbol\looparrowright 1323
\newsymbol\circeq 1324
\newsymbol\succsim 1325
\newsymbol\gtrsim 1326
\newsymbol\gtrapprox 1327
\newsymbol\multimap 1328
\newsymbol\therefore 1329
\newsymbol\because 132A
\newsymbol\doteqdot 132B
 
\newsymbol\triangleq 132C
\newsymbol\precsim 132D
\newsymbol\lesssim 132E
\newsymbol\lessapprox 132F
\newsymbol\eqslantless 1330
\newsymbol\eqslantgtr 1331
\newsymbol\curlyeqprec 1332
\newsymbol\curlyeqsucc 1333
\newsymbol\preccurlyeq 1334
\newsymbol\leqq 1335
\newsymbol\leqslant 1336
\newsymbol\lessgtr 1337
\newsymbol\backprime 1038
\newsymbol\risingdotseq 133A
\newsymbol\fallingdotseq 133B
\newsymbol\succcurlyeq 133C
\newsymbol\geqq 133D
\newsymbol\geqslant 133E
\newsymbol\gtrless 133F
\newsymbol\sqsubset 1340
\newsymbol\sqsupset 1341
\newsymbol\vartriangleright 1342
\newsymbol\vartriangleleft 1343
\newsymbol\trianglerighteq 1344
\newsymbol\trianglelefteq 1345
\newsymbol\bigstar 1046
\newsymbol\between 1347
\newsymbol\blacktriangledown 1048
\newsymbol\blacktriangleright 1349
\newsymbol\blacktriangleleft 134A
\newsymbol\vartriangle 134D
\newsymbol\blacktriangle 104E
\newsymbol\triangledown 104F
\newsymbol\eqcirc 1350
\newsymbol\lesseqgtr 1351
\newsymbol\gtreqless 1352
\newsymbol\lesseqqgtr 1353
\newsymbol\gtreqqless 1354
\newsymbol\Rrightarrow 1356
\newsymbol\Lleftarrow 1357
\newsymbol\veebar 1259
\newsymbol\barwedge 125A
\newsymbol\doublebarwedge 125B
\undefine\angle
\newsymbol\angle 105C
\newsymbol\measuredangle 105D
\newsymbol\sphericalangle 105E
\newsymbol\varpropto 135F
\newsymbol\smallsmile 1360
\newsymbol\smallfrown 1361
\newsymbol\Subset 1362
\newsymbol\Supset 1363
\newsymbol\Cup 1264
 
\newsymbol\Cap 1265
 
\newsymbol\curlywedge 1266
\newsymbol\curlyvee 1267
\newsymbol\leftthreetimes 1268
\newsymbol\rightthreetimes 1269
\newsymbol\subseteqq 136A
\newsymbol\supseteqq 136B
\newsymbol\bumpeq 136C
\newsymbol\Bumpeq 136D
\newsymbol\lll 136E
 
\newsymbol\ggg 136F
 
\newsymbol\circledS 1073
\newsymbol\pitchfork 1374
\newsymbol\dotplus 1275
\newsymbol\backsim 1376
\newsymbol\backsimeq 1377
\newsymbol\complement 107B
\newsymbol\intercal 127C
\newsymbol\circledcirc 127D
\newsymbol\circledast 127E
\newsymbol\circleddash 127F
\newsymbol\lvertneqq 2300
\newsymbol\gvertneqq 2301
\newsymbol\nleq 2302
\newsymbol\ngeq 2303
\newsymbol\nless 2304
\newsymbol\ngtr 2305
\newsymbol\nprec 2306
\newsymbol\nsucc 2307
\newsymbol\lneqq 2308
\newsymbol\gneqq 2309
\newsymbol\nleqslant 230A
\newsymbol\ngeqslant 230B
\newsymbol\lneq 230C
\newsymbol\gneq 230D
\newsymbol\npreceq 230E
\newsymbol\nsucceq 230F
\newsymbol\precnsim 2310
\newsymbol\succnsim 2311
\newsymbol\lnsim 2312
\newsymbol\gnsim 2313
\newsymbol\nleqq 2314
\newsymbol\ngeqq 2315
\newsymbol\precneqq 2316
\newsymbol\succneqq 2317
\newsymbol\precnapprox 2318
\newsymbol\succnapprox 2319
\newsymbol\lnapprox 231A
\newsymbol\gnapprox 231B
\newsymbol\nsim 231C
\newsymbol\ncong 231D
\newsymbol\diagup 231E
\newsymbol\diagdown 231F
\newsymbol\varsubsetneq 2320
\newsymbol\varsupsetneq 2321
\newsymbol\nsubseteqq 2322
\newsymbol\nsupseteqq 2323
\newsymbol\subsetneqq 2324
\newsymbol\supsetneqq 2325
\newsymbol\varsubsetneqq 2326
\newsymbol\varsupsetneqq 2327
\newsymbol\subsetneq 2328
\newsymbol\supsetneq 2329
\newsymbol\nsubseteq 232A
\newsymbol\nsupseteq 232B
\newsymbol\nparallel 232C
\newsymbol\nmid 232D
\newsymbol\nshortmid 232E
\newsymbol\nshortparallel 232F
\newsymbol\nvdash 2330
\newsymbol\nVdash 2331
\newsymbol\nvDash 2332
\newsymbol\nVDash 2333
\newsymbol\ntrianglerighteq 2334
\newsymbol\ntrianglelefteq 2335
\newsymbol\ntriangleleft 2336
\newsymbol\ntriangleright 2337
\newsymbol\nleftarrow 2338
\newsymbol\nrightarrow 2339
\newsymbol\nLeftarrow 233A
\newsymbol\nRightarrow 233B
\newsymbol\nLeftrightarrow 233C
\newsymbol\nleftrightarrow 233D
\newsymbol\divideontimes 223E
\newsymbol\varnothing 203F
\newsymbol\nexists 2040
\newsymbol\Finv 2060
\newsymbol\Game 2061
\newsymbol\mho 2066
\newsymbol\eth 2067
\newsymbol\eqsim 2368
\newsymbol\beth 2069
\newsymbol\gimel 206A
\newsymbol\daleth 206B
\newsymbol\lessdot 236C
\newsymbol\gtrdot 236D
\newsymbol\ltimes 226E
\newsymbol\rtimes 226F
\newsymbol\shortmid 2370
\newsymbol\shortparallel 2371
\newsymbol\smallsetminus 2272
\newsymbol\thicksim 2373
\newsymbol\thickapprox 2374
\newsymbol\approxeq 2375
\newsymbol\succapprox 2376
\newsymbol\precapprox 2377
\newsymbol\curvearrowleft 2378
\newsymbol\curvearrowright 2379
\newsymbol\digamma 207A
\newsymbol\varkappa 207B
\newsymbol\Bbbk 207C
\newsymbol\hslash 207D
\undefine\hbar
\newsymbol\hbar 207E
\newsymbol\backepsilon 237F
\catcode`\@=\csname pre amssym.tex at\endcsname
\def\Box{\hbox{\vrule height1ex\kern-0.4pt
\vbox to 1ex{\hrule width1ex\vfil\hrule width1ex}\kern-0.4pt\vrule height1ex}}
\newcommand{\sqr}[2]{{{\vcenter{\vbox{\hrule height.#2pt
\hbox{\vrule width.#2pt height#1pt \kern#1pt
\vrule width.#2pt}
\hrule height.#2pt}}}}}

\newcommand{\be}{\begin{equation}}

\newcommand{\ee}{\end{equation}}
\newcommand{\al}{\alpha}
\newcommand{\bt}{\beta}

\newcommand{\lm}{\lambda}

\newcommand{\rh}{\rho}

\newcommand{\phv}{\varphi}
\newcommand{\ch}{\chi}
\newcommand{\ps}{\psi}

\newcommand{\om}{\omega}

\newcommand{\raw}{\rightarrow}
\newcommand{\A}{\frak A}
\newcommand{\B}{\frak B}

\newcommand{\C}{{\Bbb C}}

\newcommand{\bib}{\bibitem}
\newcommand{\cin}{C^{\infty}}

\renewcommand{\H}{\mbox{$\cal H$}}

\newcommand{\n}{\parallel}

\newcommand{\R}{{\Bbb R}}

\newcommand{\notp}{p \kern-.48em /}
\newcommand{\ci}{\cite}
\newcommand{\bea}{\begin{eqnarray}}
\newcommand{\eea}{\end{eqnarray}}
\newcommand{\ot}{\otimes}
\newcommand{\half}{\mbox{\footnotesize $\frac{1}{2}$}}

\topmargin = - 0.5 cm
\newcommand{\PH}{{\Bbb P}{\cal H}}
\renewcommand{\P}{{\cal P}}
\newcommand{\tps}{transition probability space}
\newcommand{\Qh}{Q_{\hbar}}
\newcommand{\Ogh}{{\cal O}_{\hbar}^{\chi}}
\newcommand{\qh}{q_{\hbar}}
\newcommand{\la}{\langle}
\newcommand{\ra}{\rangle}
\newcommand{\bh}{]_{\hbar}}
\newcommand{\omh}{\omega_{\hbar}}
\begin{document}
 \setlength{\baselineskip}{1\baselineskip}
\thispagestyle{empty}
\title{Classical behaviour in quantum mechanics: a transition
probability approach\thanks{To appear in the special issue of {\em
Int. J. Mod. Phys.} {\bf B} dedicated to the memory of Hiroomi Umezawa
(1924-1995) }}
\author{ N.P. Landsman\thanks{ E.P.S.R.C. Advanced Research Fellow}\\
  Department of Applied Mathematics and Theoretical Physics\\ University of
Cambridge\\ Silver Street, Cambridge CB3 9EW, U.K. }
\date{\today}
\maketitle
\begin{abstract}
A formalism is developed for describing approximate classical behaviour in
finite (but possibly large) quantum systems. This is done in terms of
a structure common to classical and quantum mechanics, viz.\ a Poisson
space with a transition probability. Both the limit where $\hbar\raw 0$
in a fixed finite system and the limit where the size of the system goes to
infinity are incorporated. In either case, classical behaviour is seen only for
certain observables and in a restricted class of states.
\end{abstract}
\newpage
\section{Dedication}
Professor Umezawa viewed physics in a unified way based on quantum field
theory. In particular,
classical physics emerges through symmetry breaking, which leads to
condensation of Goldstone bosons
and the accompanying boson transformation \ci{UMT,Ume}.
  However, symmetry breaking only occurs in infinite systems. Even if these
exist,
it  is desirable to have a formalism that approximates the qualitative features
normally associated
with infinite systems in their finite approximants. This poses a difficult
problem for any approach
based on superselection rules (such as Umezawa's), of which formally no trace
is seen in finite
systems. Moreover, it is not
clear that all classical phenomena in Nature arise in the way described.
Indeed, examples related to
Bohr's original correspondence principle, where the classical limit arises when
certain quantum
numbers become large, do not seem to be covered.

Inspired by Umezawa's vision, we wish to present the first technical step
towards an approach to these
problems that avoids some of the difficulties mentioned. It is with sadness
that we dedicate these
pages to his memory.
 \section{Observables and states}
In quantum mechanics without superselection rules the observables  form the
self-adjoint
(Hermitian)  part of $\B(\H)$, the algebra of all bounded operators on some
Hilbert space $\H$.
In classical mechanics, the observables consist of real-valued functions on
some phase space $S$.
In both cases, they are elements of a real vector space $\A$ on which a
commutator and an
anti-commutator are defined. The former is $[A,B]_{\hbar}=i(AB-BA)/\hbar$ in
the quantum case, and is
the Poisson bracket $\{f,g\}$ in the classical case, and the latter is $A\circ
B=(AB+BA)/2$ in quantum
mechanics and $f\circ g=fg$ in classical mechanics.

The commutator satisfies the Jacobi identity, and both operations are
intertwined  by the Leibniz
rule, which says that the commutator  is a derivation of the anti-commutator.
The only difference
between classical and quantum lies in the associativity of $\circ$: classically
we have $(f\circ
g)\circ h -f\circ (g\circ h)=0$, whereas quantum-mechanically $(A\circ B)\circ
C-A\circ (B\circ C)$
equals $\hbar^2 [B,[A,C]_{\hbar}]_{\hbar}/4$. This discussion can be
generalized to systems with
superselection rules by replacing $\B(\H)$ with a general $C^*$-algebra $\A$,
and taking  a general Poisson manifold $P$ \ci{MR} rather than  the symplectic
space $S$ .

 Quantization is described by a family of maps $Q_{\hbar}:\cin(P)\raw
\A$ such that
$$ \lim_{\hbar\raw 0}( [\Qh(f),\Qh(g)\bh -\Qh(\{f,g\})) = 0 $$
 (Dirac) and  $$ \lim_{\hbar\raw 0}( \Qh(f)\circ
\Qh(g) -\Qh(fg))=0 $$ (von Neumann), for all reasonable
functions $f,g$ on $P$. There is a natural equivalence relation between
different quantizations, in
that $\Qh^1$ and $\Qh^2$ are declared equivalent if $\Qh^1(f)-\Qh^2(f)\raw 0$
for $\hbar\raw 0$ for
all $f$. This equivalence relation eliminates operator-ordering ambiguities.
 The classical limit of quantum mechanics may
be described in  similar terms, cf.\ \ci{NPL,Wer}.

While the algebraic formulation provides a nice unified description of
classical and quantum
mechanics, it is more useful for the problem at hand to give a dual description
in terms of
pure states.  Classically, we can look at $P$ as the space of pure states of
the system, which is
equipped  with a  Poisson structure. This amounts to the specification of a
Poisson bracket on a
suitable set of continuous functions on $P$, or, equivalently, may be described
in
terms of a certain geometric structure directly on $P$ \ci{MR}.

In quantum mechanics (without superselection rules to start with)  we may start
from the pure state
space $\PH$ (the projective space of the Hilbert space $\H$, obtained from the
latter by imposing $
\la\ps|\ps\ra=1$ and identifying $|\ps\ra$ with $\exp(i\al)|\psi\ra$; we denote
the image of
$|\ps\ra\in\H$ in $\PH$ by $\ps$).  There is a natural Poisson structure on
$\PH$ (which derives from
its Fubini-Study K\"{a}hler structure, cf.\ \ci{MR} or section 5 below). If we
associate a
function $f_A$ on $\PH$ to each Hermitian operator $A$ on $\H$, defined by
$f_A(\ps)=\la\ps|A|\ps\ra$, then this Poisson structure is specified by the
rule
$\{f_A,f_B\}=f_{[A,B\bh}$.

This time, however, we cannot reconstruct the system from the space $\PH$ with
its Poisson
structure, since we have not incorporated  the fact that in quantum mechanics
not all functions on
$\PH$, but only those of the form $f_A$, are observables. Also, we do not yet
know how to compute the
anti-commutator $f_A\circ f_B$. This additional information turns out to be
encoded in
the transition probabilities on $\PH$.
\section{Transition probability spaces}
A transition probability space  is a set $\P$ with a function $p$ from
$\P\times \P$ to the interval $[0,1]$, such that
$p(\ps,\phv)=1$ implies $\ps=\phv$. In standard quantum mechanics, $\P=\PH$ and
$p(\ps,\phv)=|\la\ps|\phv\ra|^2$. If there are superselection rules, the pure
state space is the union
of all sectors $\PH_S$; the transition probabilities between two different
sectors identically vanish.
 In
classical mechanics, $\P=P$ and $p^{\rm cl}(\ps,\phv)\neq 0$ only if
$\psi=\phv$ (in which case it
equals 1). Hence each point forms its own little superselection sector, cf.\
\ci{Lan0}.
See \ci{BC} for general information on \tps s.

To capture classical and quantum mechanics in one
picture, the \tps\ $\P$ should carry a unitary Poisson structure. This means
the following: given
a fixed point $\ps\in\P$, we define a function $p_{\ps}$ on $\P$ by
$p_{\ps}(\phv)=p(\ps,\phv)$.
(In standard quantum mechanics, this function is represented by the operator
$|\ps\ra\la\ps |$.)
   The
Poisson structure then leads to a vector field $\xi_{\ps}$ by the usual rule
$\xi_{\ps}f=\{p_{\ps},f\}$. This vector field defines a flow $\phv\raw \phv(t)$
for each $\phv\in\P$,
such that $d\phv(t)/dt=\xi_{\ps}(\phv(t))$. The unitarity condition is a
compatibility requirement
between the transition probabilities and the Poisson structure, viz.\ that for
every $\ps$ this flow
must leave $p$  invariant, in the sense that
$p(\phv_1(t),\phv_2(t))=p(\phv_1,\phv_2)$ for all $t$
and all $\phv_1,\phv_2\in\P$. See \ci{Lan} for details.

Given the transition probabilities, in standard quantum mechanics the Poisson
structure on $\PH$ is
actually determined by unitarity, up to a multiplicative constant, which may be
identified with
$1/\hbar$. (If there are superselection rules, each sector could in principle
have its own $\hbar$,
but this possibility does not seem to be realized in Nature.)
In classical mechanics any Poisson
structure is  unitary.

It is possible to characterize quantum mechanics (with superselection rules) in
terms of certain
axioms on a Poisson space with a transition probability, and the algebra of
observables $\A$ may
then be reconstructed  from $\P$ \ci{Lan}. An observable is here regarded as a
function on $\P$
(rather than e.g.\ an operator on a Hilbert space), and the basic point is that
every observable is a
linear combination of functions of the type $p_{\ps}$.  Moreover, every
observable
$f=\sum_i\mu_i p_{\ps_i}$ (where all $\mu_i\in\R$), has a spectral
representation
$f=\sum_{\al} \lm_{\al} p_{e_{\al}}$, where
$p(e_{\al},e_{\bt})=\delta_{\al\bt}$.
This spectral theorem holds for more general \tps s than those describing
quantum mechanics;
even in the latter case one is led to a new proof of the usual spectral theorem
(which is a
restatement of the one used here), which does not use Hilbert space theory
\ci{Lan}.

The spectral representation   is used to define the anti-commutator via
$f^2\equiv f\circ f=\sum_{\al} \lm_{\al}^2 p_{e_{\al}}$, and then $f\circ g=
((f+g)^2-(f-g)^2)/4$. Unitarity then implies that the commutator (defined as
the Poisson
bracket) and the anti-commutator are related by the Leibniz identity.
The reader may immediately check that in classical mechanics (where the sums
extend over an
uncountable number of terms) every observable is automatically in spectral
form, so that
 $f\circ g=fg$ in the usual sense of pointwise multiplication. In quantum
mechanics, however, one
recovers the usual anti-commutator in the above way.
Finally, the norm of an observable is simply given by $\n f\n={\rm
sup}_{\ps\in\P}|f(\ps)|$.
\section{Classical germs}
{}From the point of view of Poisson spaces with a transition probability,
quantization theory and the
classical limit of quantum mechanics are described in one and the same way.
One starts from  a Poisson manifold $P$ (the pure state space of the classical
system) and a quantum
pure state space $\P$ (e.g., $\P=\PH$ for some Hilbert space $\H$). The basic
ingredient is a family
of injections $\qh:P\raw \P$  (defined for $\hbar$ in a certain interval
$(0,\hbar_0$)) which satisfy
$$ \lim_{\hbar\raw 0} p(\qh(x),\qh(y))= \delta_{xy}$$ for  all $x,y\in P$
(recall
that  $\delta_{xy}$ is just the classical transition probability $p^{\rm
cl}(x,y)$).
Thus each $\qh$ embeds the classical state space into its quantum counterpart,
in such a way that
for small $\hbar$ the classical points all become almost mutually orthogonal.
 Such a family $\qh$ generalizes the notion of a coherent state, and may be
referred to as a
classical germ \ci{NPL} (also cf.\ \ci{Wer}). Two germs $\qh^1$ and $\qh^2$ may
be declared
equivalent if $\lim_{\hbar\raw 0}p(\qh^1(x),p(\qh^2(x))=1$ for all $x\in P$.

 Consider a function  $f$ on $P$ which is nonzero only at a finite number of
points; its spectral
representation is $f= \sum_x f(x) p^{\rm cl}_x$. Using a classical germ, we can
define a
quantization of $f$ by $\Qh(f)=\sum_x f(x) p_{\qh(x)}$.
For small $\hbar$ the r.h.s.\  will approximate the spectral representation of
$\Qh(f)$, so that
$\Qh(f)^2$ is approximately $\sum_x f(x)^2 p_{\qh(x)}$, which equals
$\Qh(f^2)$.
Hence $\Qh(f)^2\raw \Qh(f^2)$ for $\hbar\raw 0$, which if true for all $f$ is
equivalent to
von Neumann's condition.
In practice, $f$ will have support in an
uncountable set, and the sum will be replaced by an integral. The above
prescription then suggests
quantizations of the type $\Qh(f)=\int_P d\mu(x)\, f(x)  p_{\qh(x)}$, where the
measure $\mu$ is
normalized by the requirement $\Qh(1)=1$, and, in case that $P$ is symplectic,
is usually
proportional to the Liouville measure.  Since in usual notation $
p_{\qh(x)}=|\qh(x)\ra\la \qh(x)|$,
we see that coherent state quantization schemes (cf.\ \ci{Per}) are a special
case of this.

Given an equivalence class of classical germs, one can consider a family
$A_{\hbar}$ of observables on
$\P$ (we regard an observable as a function on $\P$, and will not distinguish
bewteen an operator
$A$ and its associated function $f_A$)  which depend on $\hbar$ in such a way
that $\lim_{\hbar\raw 0}
A_{\hbar}(\qh(x))\equiv A_0(x)$ exists for all $x$, and defines a continuous
function $A_0$ on $P$.
 (In
that case, the family $A_{\hbar}$ may be seen as a quantization of $A_0$.)
We will refer to such a family as a classical funnel.

For any function $F$ on $\P$, denote by $\qh^*F$ the function on $P$ defined by
$\qh^*F(x)=F(\qh(x))$. We then impose the requirement on the classical germ
that
for those classical funnels $A_{\hbar}$ and $B_{\hbar}$ for which the limit
functions $A_0$ and
$B_0$ are differentiable, $\qh^*[A_{\hbar},B_{\hbar}\bh$ approaches the Poisson
bracket $ \{
\qh^*A_{\hbar},\qh^*B_{\hbar}\}$ when $\hbar\raw 0$. Cf.\ \ci{Yaf} for the
coherent state analogue,
and \ci{Wer} for necessary conditions on $\qh$.  This requirement is the state
space analogue
 of Dirac's condition.
\section{Classical germs from coherent states}
An interesting class of examples of classical germs comes from a particular
type of coherent states,
which we will now describe in a geometric way. The material in  the next
paragraph may be
found in \ci{MR}, and a heuristic presentation is in \ci{Yaf}.

Any Hilbert space $\H^{\ch}$ carries a canonical symplectic form $\om_{\hbar}$.
If we identify the
tangent space $T\H^{\ch}$ with $\H^{\ch}$, this is defined by
$\omh(|\ps\ra,|\phv\ra)=-2\hbar \,{\rm
Im}\,\la\phv|\ps\ra$ (we assume $\hbar\neq 0$). It quotients to the projective
space $\PH^{\ch}$, on which
it gives the Fubini-Study form.  Now let $U^{\ch}$ be an irreducible  unitary
representation of a connected Lie group $G$ on $\H^{\ch}$.   This naturally
defines an action of $G$
(which we denote by the same symbol $U^{\ch}$) on
$\PH^{\ch}$, which turns out to be strongly Hamiltonian. Thus we find an
equivariant momentum map
$J^{\ch}_{\hbar}:\PH^{\ch}\raw {\frak g}^*$, where $\frak g$ is the Lie algebra
of $G$, and ${\frak g}^*$
its dual.  This just means that for any generator $T_a$ of $\frak g$ we have a
function
$<J^{\ch}_{\hbar},T_a> \equiv J_a$ on
$\PH^{\ch}$ which generates the action of $G$ on $\PH^{\ch}$ as a canonical
transformation.
 The Poisson brackets
of the $J_a$ reproduce the Lie algebra , i.e., $\{J_a,J_b\}=f_{ab}^cJ_c$.
This is equivalent to global equivariance, that is, $J^{\ch}_{\hbar}\circ
U^{\ch}={\rm Co}\circ
J^{\ch}_{\hbar}$, where Co is the co-adjoint action of $G$ on ${\frak g}^*$.
Explicitly,
$J_a(\ps)=-i\hbar \la\ps|dU^{\ch}(T_a)|\ps\ra$, where $\la\ps|\ps\ra=1$, and
 $dU^{\ch}(T_a)$ is the anti-Hermitian representative of $T_a$ (so that
$[dU^{\ch}(T_a),dU^{\ch}(T_b)]=f_{ab}^c\, dU^{\ch}(T_c)]$).

Take a fixed $\ps_0\in\PH^{\ch}$. It may happen that its orbit
$U^{\ch}(G)\ps_0$ is a symplectic
subspace of $\PH^{\ch}$. In that case, $U^{\ch}(G)\ps_0$ is a covering space of
the co-adjoint orbit $\Ogh$
through $J(\ps_0)$ in ${\frak g}^*$ (see Thm.\ 14.6.5 in \ci{MR}; $\Ogh$ is
here assumed to be
equipped with its canonical symplectic form).
 In most examples,  $U^{\ch}(G)\ps_0$ is actually homeomorphic to $\Ogh$; the
momentum
map then provides an identification of the two as symplectic spaces.
The dependence of $\Ogh$ on $\hbar$ comes from the fact that the symplectic
form on $\Ogh$ is
proportional to $\hbar$. In cases of interest to the classical limit of quantum
mechanics,
the label $\ch$ is of the form $\ch=L\ch_0$, where $L$ is some positive number
(which is quantized if
$G$ is compact, as is $\ch_0$). In that case, the orbit $\Ogh$ coincides with
 ${\cal O}^{\ch_0}_1\equiv {\cal O}^{\ch_0}$ as
a manifold, and has the symplectic form $\hbar L\, \om_{\ch_0}$, where
$\om_{\ch_0}$ is the symplectic
form on ${\cal O}^{\ch_0}$. In the classical regime in the sense of Bohr, the
quantum number $L$ is
very large. With $\hbar$ a fixed constant of Nature, this means that $\Ogh$
will blow up as
$L\raw\infty$. To avoid this, in practice one keeps the classical scale $\hbar
L$ fixed (and equal
to 1), and stipulates  that this fixed scale is large compared to $\hbar$. This
 means
that one lets $\hbar\raw 0$ and therefore (still) $L\raw\infty$. If $L$ is
quantized then clearly
$\hbar$ can no longer assume arbitrary  values, and the classical limit is
achieved   along a
sequence $\{\hbar=1/L\}_{L\in{\Bbb N}}$. This latter procedure is the one we
will follow. In
particular, $\ch$ blows up in the classical limit. If necessary, the classical
scale may be varied by
changing $\ch_0$.

For fixed $\hbar>0$, we define $\qh(x)=  (J^{\ch}_{\hbar})^{-1}(x)$.
If $x={\rm Co}\,(g) J(\ps_0)$ then $\qh(x)$ corresponds to the coherent state
$U^{\ch}(g)|\ps_0\ra$ in
the usual formalism \ci{Per}. In view of the above, we see that the states
$\qh(x)$  lie in different
(projective) Hilbert spaces as $\hbar$ varies. This is no problem in the
context of our formalism, as
we may take the space of pure states $\P$ to be  the union of all projective
unitary representation
spaces of $G$ (this is nothing but the pure state space of the group algebra
$C^*(G)$, and therefore
a perfectly natural object). We then look at the $\qh$ as a collection of
injective maps from ${\cal
O}^{\ch_0}$ into $\P$. It may then be verified that the $\qh$ indeed define a
classical germ.
For a given value of $\hbar$, the lack of classical behaviour is measured by
the
non-zero-ness of the transition probabilities $p(\qh(x),\qh(y))$ for $x\neq y$.

Classical funnels $A_{\hbar}$ are functions of the operators $-i\hbar
dU^{\ch}(T_a)$;
that is, $A_{\hbar}$ is nonzero only on $\PH^{\ch_0/\hbar}$. If
$A_{\hbar}=a(-i\hbar
dU^{\ch}(T_1),\ldots, -i\hbar dU^{\ch}(T_n))$ for some function $a$, then
$A_0=a(T_1,\ldots,T_n)$,
regarded as a function on ${\cal O}^{\ch_0}$, cf.\ \ci{Yaf}.
\section{Coherent state examples of classical  germs}
The following examples are well known (e.g., \ci{Yaf,Bon1,ZF}), but it is
useful to see them
reformulated in the language described above.

Consider the Heisenberg group $G=H_n$ in $n$ dimensions. The generators of
${\frak g}\simeq G \simeq
\R^{2n+1}$ are $\{P_i,Q_j,Z\}$ ($i,j=1\ldots n$), with Lie brackets
$[P_i,Q_j]=\delta_{ij}Z$
and $[P_i,Z]=[Q_j,Z]=0$. We denote the dual basis in ${\frak g}^*$ by
$\{\hat{P}_i,\hat{Q}_j,\hat{Z}\}$.
The co-adjoint orbit ${\cal O}^{\lm}$ ($\lm\neq 0$) through $\lm\hat{Z}$ may be
identified with
$T^*\R^n$; under this identification, a point $
({\bf p}, {\bf q})\in T^*\R^n$ is identified with $ p_i\hat{P}_i+
q_j\hat{Q}_j+\lm\hat{Z}$ (summation
convention). One finds $${\rm Co}(p_jQ_j-q_iP_i)\lm\hat{Z}=p_i\hat{P}_i+
q_j\hat{Q}_j+\lm\hat{Z}.$$
 The  symplectic form on ${\cal O}^{\lm}$ is given by
$\lm dp_i\wedge dq_i$.

The irreducible representations $U^L$ ($L\neq 0$) are all realized
on the same Hilbert space
$\H^L\equiv \H=L^2(\R^n)$, and may be specified by $dU^L(P_i)=
\partial/\partial x_i$,
$dU^L(Q_j)=iLx_j$, and $dU^L(Z)=iL$. If we now take $\ps_0\in\PH$  defined by
$$\la
x|\ps_0\ra=(L/\pi)^{n/4}\exp(-Lx^2/2)$$ then $J_{\hbar}^L(\ps_0)= \hbar
L\hat{Z}$.
The classical germ $\qh$ is   a family of maps from ${\cal O}^1\simeq T^*\R^n$
into $\PH$. Hence we put $L=1/\hbar$. By construction, $\qh ({\bf p}, {\bf q})$
is then  given
by
$U^{1/\hbar}(p_jQ_j-q_iP_i)\ps_0\ $, which is represented in $\H$ by the wave
function  $$\la x|\qh({\bf p}, {\bf q})\ra= (\pi\hbar)^{-n/4}\exp(-\half i
p_jq_j/\hbar)
\exp(ip_jx_j/\hbar)\exp(-(x-q)^2/2\hbar).$$ One checks without difficulty that
all requirements on a
classical germ are indeed satsified. Classical operators are functions of
$-i\hbar
dU^{1/\hbar}(T_a)$, where $T_a$ is $P_i$, $Q_j$, or $Z$. Clearly, $$-i\hbar
dU^{1/\hbar}(P_i)=
-i\hbar \partial/\partial x_i; \:\: -i\hbar dU^{1/\hbar}(Q_j)=x_j; \:\: -i\hbar
dU^{1/\hbar}(Z)=1.$$

In the next round of examples, $G$ is a connected compact semi-simple Lie
group.
What follows is merely a reformulation of some of the results in
\ci{Per,Ono,Yaf,Sim,Duf}.
The label $\ch$ stands for a highest weight (relative to a choice of a maximal
torus $T\subset G$ and
of a fundamental Weyl chamber), and we assume that  $\ch=L\ch_0$ for $L\in
{\Bbb N}$ and some highest
weight $\ch_0$. Each  such $\ch_0$ defines a co-adjoint orbit ${\cal
O}^{\ch_0}$; this is the
orbit through $\ch_0$, which originally was an element of ${\frak t}^*$ but is
now regarded as an
element of ${\frak g}^*$ by putting it equal to zero on the orthocomplement of
${\frak t}$ in $\frak
g$ with respect to the Killing metric. If $|\Psi_0\ra$ is the highest weight
vector in $\H^{\ch}$,
then $J_{\hbar}^{\ch}\Psi_0$ lies in the orbit through $\hbar L\ch_0$, and
$J_{\hbar}^{\ch}$ is a
symplectomorphism between $U^{\ch}(G)\Psi_0$ and ${\cal O}^{\hbar \ch}$. As
explained
above, we now take ${\cal O}^{\ch_0}$ as the fixed classical phase space, and
put $\hbar=1/L$.
The map $\qh$ then injects ${\cal O}^{\ch_0}$ into $\PH^{\ch}$, and defines a
classical germ.

For example, for $G=SU(2)$ the co-adjoint orbit ${\cal O}^{j_0}$ is a sphere
$S^2$ in ${\frak
g}^*=\R^3$ with radius $j_0$; the symplectic form is  $j_0$ times the
Fubini-Study form on $S^2\simeq
{\Bbb P}\C^2$.  With $j_0=1$ and $\hbar=1/j$, one finds that
$p(\qh(z),\qh(w))=(\cos\half\Theta)^{4/\hbar}$, where $\Theta$ is the angle
between $z$ and $w$.
Clearly, $\lim_{\hbar\raw 0} p(\qh(z),\qh(w))=\delta_{zw}$.

Our last example of this sort (cf.\ \ci{ZF}) provides a bridge towards the
systems studied in the next
section.
 We take $G=U(M)$; its Lie algebra comprises the set of observables of an
$M$-level system.
 We realize ${\frak g}^*$ as the space of all Hermitian
$M\times M$ matrices $\rh$, with pairing $<\rh,T_a>=-i{\rm Tr}\, \rh
T_a$, where
 $T_a\in{\frak g}$
is realized  in its defining representation (i.e., as a skew-Hermitian $M\times
M$ matrix).
A co-adjoint orbit is labeled by an $M$-tuple of real numbers, and consists of
all Hermitian matrices
having these numbers as  eigenvalues. The co-adjoint orbit ${\cal O}^1$ of
interest
equals the set of matrices $\rh$ with eigenvalues $(1,0,\ldots,0)$, and
corresponds to the highest weight
$\ch_0=(1,0,\ldots,0)$. Any such $\rh$ can be writen as $\rh=|\ps\ra\la\ps|$
for some $|\ps\ra\in
\C^M$  with $\la\ps|\ps\ra=1$. It follows that ${\cal O}^1\simeq {\Bbb P}\C^M$,
equipped with the
Fubini-Study symplectic form. When convenient, we label its points simply by
$\ps$, which
of course stands for the matrix $|\ps\ra\la\ps|$. This is our fixed classical
phase space. The highest
weight $\ch_0$, in turn, corresponds to  the defining representation
$U^{\ch_0}\equiv U^1$ of $U(M)$
on $\H^{\ch_0}\equiv \H^1=\C^M$.

The representation $U^{L\ch_0}\equiv U_S^L$ is realized on
$\H_S^{L}=\otimes^L_S\C^M$, the symmetrized tensor product of $L$ copies of
$\C^M$. This is the
state space of a system of $L$ identical bosons.
The momentum map in this representation is  given by linear extension of
%% FOLLOWING LINE CANNOT BE BROKEN BEFORE 80 CHAR
$$J_{\hbar}^L(\ps_1\otimes_S\ldots\otimes_S\ps_L)=\hbar\sum_{i=1}^L|\ps_i\ra\la\ps_i|.$$
The highest weight vector $|\Psi_0\ra$ in $\H_S^L$ is the $L$-fold tensor
product
$\otimes^L |{\bf e}_1\ra$ of $L$
copies of the first  basis vector $|{\bf e}_1\ra$. For the coherent states we
get
 $J_{\hbar}^L(\otimes^L \ps)=\hbar L |\ps\ra\la\ps| $. (Since
$U^L_S(g)\Psi_0=\otimes^L U(g){\bf
e}_1$,   any point in the orbit $U_S^L(G)\Psi_0$ has the   form $\otimes^L
\ps$.)
 Thus $J_{\hbar}^L$ provides
a symplectomorphism between $U_S^L(G)\Psi_0$ and the co-adjoint orbit
consisting of all matrices with
eigenvalues $(\hbar L,0,\ldots,0)$. As before, we now put $\hbar=1/L$. By the
general theory, this
leads to a classical germ, whose member $\qh$ injects ${\cal O}^1$ into
$\H_S^L$. Explicitly,
$\qh(\ps)=\otimes^L\ps$.  Hence
$$p(\qh(\ps),\qh(\phv))=|\la\phv|\ps\ra|^{2L},$$ and it immediately
follows that $\lim_{\hbar\raw 0}p(\qh(\ps),\qh(\phv))=\delta_{\ps\phv}$.

Finally, we note that classical funnels  must be functions of $-i \hbar
dU_S^L(T_a)=
(-i/L)\sum_{j=1}^L T_a^{(j)}$, where $T_a^{(j)}$ is defined by linear extension
of
the operator
$$T_a^{(j)}|\ps_1\ra\otimes_S\ldots\otimes_S|\ps_L\ra= |\ps_1\ra\otimes_S\ldots
 T_a|\ps_j\ra \ldots \otimes_S|\ps_L\ra.$$ The corresponding function
$f_{T_a^{(j)}}$ on $\PH_S^L$ is
simply given by
$$f_{T_a^{(j)}}(\ps_1\otimes_S\ldots\otimes_S\ps_L)=f_{T_a}(\ps_j).$$
 Thus a classical funnel must be a function of single-particle operators
averaged over all bosons
in the system. This is a consequence of the irreducibility of $U_S^L$, which
would not hold if the
particles were distinguishable (so that the tensor product $\otimes^L\C^M$ is
not symmetrized).
\section{Mean-field systems}
A  spectacular occurrence of classical behaviour in a quantum system is
encountered in mean-field
systems. These include certain formulations of the BCS model of
superconductivity, the Dicke laser
model, Josephson junctions, etc. The current theoretical understanding of these
models has emerged
from the papers \ci{HL,MS,Bon1,Bon2,DR,RW1,Unn,DW}, and others. Below we only
study the so-called
homogeneous case.

As in the previous section, we look at  $L$ copies of an $M$-level system, but
this time the
partcicles are distinguishable, and may be thought of as sitting at the points
of a lattice
containing $L$ sites.
  The algebra of observables $\A_{\infty}$ of the infinite
system ($L=\infty$) is defined as the ($C^*$-inductive) limit of the algebra
generated by operators
of the type $A_1\otimes A_2\ldots \otimes A_N\otimes 1\ldots$, where $N$ is
finite (but varies), and
the tail only consists of unit operators 1. Classical behaviour is found by
focusing on the subset
of the pure state space consisting of the   permutation invariant states (such
states are invariant
under permutations of the $A_i$ in the string above). It can be shown \ci{Sto}
that  permutation
invariant pure states are of the form $ \otimes^{\infty}\ps\equiv\Psi$, where
$\ps$ is a pure state on
$\A_1$, i.e., $\ps\in {\cal O}^1={\Bbb P}\C^M$. Hence the pure permutation
invariant states
themselves form the space ${\Bbb P}\C^M$. The transition
probability between different $\Psi$'s vanishes - this is intuitively obvious
from the previous
section, since the permutation invariant states act on the observables as if
the particles were
indistinguishable bosons. Moreover, the transition probability between
arbitrary  local
perturbations of $\Psi$ and $\Psi'\neq \Psi$ vanishes.  (In fact, the subset
$S^P$ of all permutation
invariant states of the total state space of $\A_{\infty}$ is a so-called Bauer
simplex, whose
boundary consists of the primary, or `macroscopically pure',  permutation
invariant states.  Hence
all states in $S^P$ have a unique decomposition into states which describe pure
phases of the system.
These properties make $S^P$ a classical object.)

We now return to the finite system ($L<\infty$), and take the pure state space
$\P$ to be the union
of all $\P_L$, where $\P_L$ is the pure state space of $\A_L$ (the $L$-particle
system).
Clearly, $\P_L=\PH^L$, with $\H^L=\ot^L\C^M$.  The group $G=U(M)$ acts on
$\H^L$ by the
reducible unitary representation $U^L=\ot^L U^1$. The corresponding momentum
map is given by
essentially the same formula as in the symmetrized case.
 With $\hbar=1/L$ in what follows, we are therefore led to define the classical
germ as a family of
maps from ${\cal O}^1$ (as in the previous example) into $\P$. In view of the
above, we take
$\qh(\ps)=\otimes^L \ps\in \P_L\subset \P$.  The transition probability is
$p(\qh(\ps),\qh(\phv))=|\la\phv|\ps\ra|^{2L}$, exactly as before.

The preceding paragraphs relate to any lattice model. What characterizes
homogeneous mean-field
models is their time-evolution. Namely, the Hamiltonian $H_L$ of the
$L$-particle system is assumed to
be of the form $H_L=L \tilde{H}$ where   $\tilde{H}$
is a function of the scaled generators  $-i \hbar dU^L(T_a)$ of $U(M)$ (cf.\
\ci{DW} for a wider
class of Hamiltonians). Here
 $-i \hbar dU^L(T_a)=
(-i/L)\sum_{j=1}^L T_a^{(j)}$, where  $T_a^{(j)}$ is given by essentially the
same expression as
in  the symmetrized case, i.e., it acts as $T_a$ on the $j$'th copy of $\C^M$.
If  $\tilde{H}$ is
non-linear, a particle interacts with all other particles. Note that there are
many more classical
funnels than those of the type  $\tilde{H}$ alone: for example, all operators
in $\A_{\infty}$ are
included.

In view of the long-range nature of the Hamiltonian, a time-evolution on the
infinite system
$\A_{\infty}$ does not exist in the usual sense (that is, as a one-parameter
automorphism group of
the algebra of observables). Instead, the limit $L\raw\infty$ of the evolution
described by $H_L$
only exists in certain representations. These include those induced by
permutation  invariant states.
For those states (and their local perturbations) one can define a limiting
Schr\"{o}dinger picture
time-evolution.

Recall that the
permutation invariant pure states $\Psi$ of the infinite system form the
manifold ${\Bbb P}\C^M$,
which we identify with the  co-adjoint orbit ${\cal O}^1$ in ${\frak g}^*$,
equipped with its
canonical symplectic structure.
 The spectacular fact is now that the full quantum
Schr\"{o}dinger time-evolution of the states $\Psi$ in  ${\cal O}^1$  coincides
with the classical
time-evolution
generated by $\tilde{H}$, now regarded as a function on  ${\cal O}^1$
through the replacement of   $-i \hbar dU^L(T_a)$ by $T_a$ in its arguments
(here $T_a\in{\frak g}$
is seen as a function on ${\frak g}^*$, and hence on its subspace  ${\cal
O}^1$, by linear
evaluation).
In particular, quantum ground states simply correspond to stationary points of
the classical
dynamics.

For the finite systems this means, from our point of view, that
$$\lim_{\hbar\raw 0}p(\qh(\ps)(t),\qh(\ps(t)))=1,$$ where $\ps\raw \ps(t)$ is
the classical
time-evolution in ${\cal O}^1$ generated by $\tilde{H}$, and $
\qh(\ps)\raw\qh(\ps)(t)$ is the
time-evolution in $\P_L$ (with $L=1/\hbar$, as always) generated by the
Hamiltonian $H_L$.
 For each fixed finite size $L$ we can monitor the departure from classical
behaviour
by computing the static transition probabilities $p(\qh(\ps),\qh(\phv))$ and
the dynamical ones $p(\qh(\ps)(t),\qh(\ps(t)))$.
The hypothetical infinite system  only enters through its classical shadow, the
finite
phase space ${\cal O}^1$.

In our opinion, these models provide strong support for the belief that the
existence of the
classical world is compatible with quantum mechanics.

\end{document}